\documentclass[copyright,creativecommons]{eptcs}
\usepackage{breakurl}             

\usepackage{verbatim}
\usepackage{tikz}
\usepackage{color}
\usepackage{hyperref}
\usepackage{pstricks,pst-node,pst-tree}
\usetikzlibrary{positioning}
\usetikzlibrary{shapes.geometric}
\usetikzlibrary{shapes.misc}
\tikzstyle{boite}=[rectangle,rounded corners,fill=red!30,draw=red!50,text centered, minimum height=1cm,top color=white, bottom color=red!50!black!20,very thick]
\tikzstyle{boiterect}=[rectangle,fill=gray!30,draw=gray!50,text centered, minimum height=1cm,top color=white, bottom color=gray!70!black!20,very thick]
\tikzstyle{boiteround}=[rounded rectangle,fill=blue!30,draw=blue!50,text centered, minimum height=1cm,top color=white, bottom color=blue!70!black!20,very thick]
\tikzstyle{inout}=[cylinder, shape border rotate=90, fill=red!30,draw=red!50,aspect=0.25,minimum height=1.5cm,text centered,top color=white, bottom color=red!70!black!20,very thick,text width=1.5cm]

\title{Solving Linux Upgradeability Problems\\ Using Boolean Optimization}
\author{Josep Argelich 
\institute{DIEI}
\institute{Universitat de Lleida}
\email{jargelich@diei.udl.cat}
\and Daniel {Le Berre} 
\institute{CRIL-CNRS UMR 8188}
\institute{Universit\'e Lille-Nord de France}
\email{leberre@cril.fr}
\and In\^es Lynce
\institute{INESC-ID/IST} 
\institute{Technical University of Lisbon, Portugal}
\email{ines@sat.inesc-id.pt}
\and Joao Marques-Silva 
\institute{CSI/CASL}
\institute{University College Dublin, Ireland}
\email{jpms@ucd.ie}
\and Pascal Rapicault
\institute{Sonatype}
\institute{Mountain View, CA 94040, USA}
\email{pascal@sonatype.com}
}

\def\titlerunning{Solving Linux Upgradeability Problems Using Boolean Optimization}
\def\authorrunning{J. Argelich, D. {Le Berre}, I. Lynce, J. Marques-Silva \& P.  Rapicault}

\newtheorem{definition}{Definition}

\begin{document}
\maketitle

\begin{abstract}
Managing the software complexity of package-based systems can be regarded as one of the main challenges in software architectures. Upgrades are required on a short time basis and systems are expected to be reliable and consistent after that. For each package in the system, a set of dependencies and a set of conflicts have to be taken into account. Although this problem is computationally hard to solve, efficient tools are required. In the best scenario, the solutions provided should also be optimal in order to better fulfill users requirements and expectations.
This paper describes two different tools, both based on Boolean satisfiability (SAT), for solving Linux upgradeability problems. The problem instances used in the evaluation of these tools were mainly obtained from real environments, and are subject to two different lexicographic optimization criteria. The developed tools can provide optimal solutions for many of the instances, but a few challenges remain.
Moreover, it is our understanding that this problem has many similarities with other configuration problems, and therefore the same techniques can be used in other domains.
\end{abstract}

\section{Introduction}

One of the current challenges in open source software distributions, such as Linux distributions, consists in managing the complexity of software package upgrades. Typically, each package is associated with a set of package dependencies and a set of package conflicts. For a package to be installed, the depending packages must be installed as well, whereas the conflicting packages cannot be installed. It is straightforward to realize that such implications may easily lead to a chain of packages to be installed and removed. In a system with thousands of packages, the request for installing, upgrading or removing a single package may have a huge impact on the system and create a non-trivial problem to be solved. 

To solve this problem, a few alternative approaches have been proposed and integrated in Linux distributions. However, these solutions are usually not complete, as the problem may have a solution, which is not guaranteed to be found even if given enough resources. The use of heuristic approaches is perfectly justified if one considers that this problem is NP-hard~\cite{MBCVDLT06}. Still, it is worth investigating optimal solutions. State-of-the-art Boolean satisfiability and optimization tools are currently able to solve huge and difficult problems, which justifies evaluating these tools in the context of the package upgradeability problem.

This paper describes Boolean-based tools for solving Linux dependencies. These tools have been developed in the context of the Mancoosi internal solver competition~\footnote{\url{http://www.mancoosi.org/misc-live/}}, where an input format is pre-defined, as well as a couple of optimization criteria. The proposed solutions are guaranteed to find an optimal solution if the required resources are available. In most of the cases, an optimal solution is found.

The remaining of the paper is organized as follows. The next section introduces fundamental concepts related with Boolean satisfiability and Boolean optimization. Section~\ref{sec:dep} describes the upgradeability problem and explains the main rules of the Mancoosi competition. The next section describes the P2CUDF tool that solves upgradeability problems using Pseudo-Boolean Optimization (PBO). The next section describes the INESC-ID tool that solves upgradeability problems using Maximum Satisfiability (MaxSAT) while taking advantage of P2CUDF as a front-end. Finally, experimental results are provided and the paper concludes.  

\section{Boolean Satisfiability and Optimization}

A Boolean formula is often assumed to be in the Conjunctive Normal Form (CNF), where a formula is a conjunction of clauses, a clause is a disjunction of literals and a literal is a Boolean variable (i.e. a positive literal) or its negation (i.e. a negative literal). 

A Boolean variable may be assigned truth values {\tt true} or {\tt false} (or values 1 or 0, respectively). Given a (partial) assignment to the Boolean variables in a Boolean formula, the formula is satisfied {\em iff} all of its clauses are satisfied and a clause is satisfied {\em iff} at least one of its literals is satisfied. A positive literal is satisfied {\em iff} the corresponding variable is assigned value {\tt true} and a negative literal is satisfied {\em iff} the corresponding variable is assigned value {\tt false}.

The problem of deciding whether a Boolean formula is satisfiable, i.e. if there exists an assignment to the variables such that the formula is satisfied, is called Boolean satisfiability (SAT). SAT was the first problem that was proved to be NP-complete~\cite{cook-stc71}, and has motivated theoretical investigation in the field. Furthermore, SAT finds many practical applications such as hardware and software verification, planning, cryptography and bioinformatics. Modern SAT solvers implement backtrack search enhanced with powerful reasoning techniques, accurate heuristics and dedicated data structures.

Given a Boolean formula, the MaxSAT problem is the optimization problem to determine the largest number of clauses that can be satisfied. This problem can also be seen in its MinUNSAT version, which is determining the smallest number of clauses that cannot be satisfied. MaxSAT can be seen as a generalization of SAT: if the solution to the MaxSAT problem is the total number of clauses in the formula, then that formula is SAT, else it is UNSAT. The MaxSAT problem has a few interesting variants, namely Partial MaxSAT, Weighted MaxSAT and Partial Weighted MaxSAT. In Partial MaxSAT some clauses are classified as {\em hard} and {\em must} be satisfied, whereas the remaining ones are classified as {\em soft} and {\em should} be satisfied (in the MaxSAT problem, all the clauses are {\em soft}). A solution to the Partial MaxSAT problem satisfies all the hard clauses and maximizes the number of satisfied soft clauses. In Weighted MaxSAT each clause is soft and associated to a weight, and the goal is to maximize the sum of the weights of satisfied clauses. Partial Weighted MaxSAT allows hard clauses in the latter problem.

In a pseudo-Boolean formula, variables have Boolean domains and constraints (called PB-constraints) are linear inequalities with integral coefficients. In Pseudo-Boolean optimization (PBO), a cost function is added to a Pseudo-Boolean formula.  

A detailed description of modern SAT solvers, maximum satisfiability and Pseudo-Boolean optimization can be found, respectively, in~\cite{jpms-sathbk09,manya-sathbk09,manquinho-sathbk09}.

\section{The upgradeability problem}
\label{sec:dep}

The basic definitions for the upgradeability problem are given below. These definitions closely follow the ones provided in~\cite{TSJL07}.

\begin{definition} (Universe)
A universe $U$ is a finite set of package rules, where each package rule is a tuple of the form $(p,D,C)$, where $p$ is a package characterized by its name and version, and:
\begin{itemize} 
\item D is a set of dependency clauses for p that stipulate which packages must be present in order to install the package $p$. Each dependency clause is a disjunction of packages $p_1 \vee \ldots \vee p_k$, which means that at least one of the packages $p_1, \ldots, p_k$ must be installed for the package $p$ to work properly. 
\item C is a set of conflict clauses for $p$ that stipulate which packages must not be installed for the package $p$ to work properly.
\end{itemize} 
\end{definition}

\begin{definition} (Valid Installation Profile)
An installation profile for a universe is a subset of the packages of the universe. An installation profile is said to be valid if the requirements for each one of its packages are satisfied, i.e. if dependency and conflict clauses are satisfied for each package.
\end{definition} 

\begin{definition} (Upgradeability Problem)
Given a universe $U$, an installation profile $P$, and a new package $p$, the upgradeability problem consists in deciding whether there exist a set of packages $P^+$ containing $p$ and a set of packages $P^-$ not containing $p$ such that $P \cup P^+ - P^-$ is a valid installation profile for $U$. 
\end{definition} 

Observe that a solution to the upgradeability problem considers a set of packages $P^+$ (which must include the new package $p$) to be added to the set of already installed packages $P$ as well as a set of packages $P^-$ (which must not include the new package $p$) to be removed from $P$.

Roughly speaking, solving the upgradeability problem is the target of the Mancoosi project. Mancoosi~\footnote{\url{http.//www.mancoosi.org}} is a research project funded by the European Commission in the 7th Framework Programme (FP7). The project involves 10 academic and industrial partners. The main goal of Mancoosi is to develop better tools for package-based system administration, which is divided into two main sub-goals: (i) provide support for rollback of package installations and (ii) provide better tools for the resolution of dependencies and conflicts between packages.

In the context of the second sub-goal, a competition for package installation tools has been organized. In a few words, the tools to be developed are required to be (i) complete, (ii) powerful and (iii) efficient. In detail, these tools should be able to (i) find a solution whenever there exists one, (ii) understand complex optimization criteria, and (iii) find a good solution fast, despite of the theoretically high complexity of the problem. 

The tools entering the competition are required to have a common format for inputs and results, which has been called Common Upgradability Description Format (CUDF). CUDF combines features of the RPM and the Debian packaging world, and also allows to encode other formats like for instance metadata of Eclipse plugins. However, the format has been simplified in order to make life easier for the development of problem solvers. 

Given a user request for installing one specific package, two optimization criteria have been proposed. Both are lexicographic combinations of some simple integer valued utility functions of a solution. The first one is called {\em paranoid} and answers the user request, minimizing the number of packages removed in the solution, and also the packages changed by the solution. The second one is called {\em trendy} and answers the user request, minimizing the number of packages removed in the solution, minimizing the number of outdated packages in the solution and finally minimizing the number of extra packages installed\footnote{The trendy criterion has been recently updated with the criteria {\em minimizing the number of package recommendations that are not satisfied} inserted between the second and third criteria.}. 

\section{P2CUDF: Solving CUDF through Pseudo-Boolean Optimization}
\begin{figure}[ht]
\begin{center}
\begin{tikzpicture}
\node (cudf)[inout] at (-0.6,2) {CUDF Request};
\node (cudfsol)[inout] at (12,2) {CUDF Solution};

    \path  [fill=orange!10, dashed, draw=black, rounded corners] 
    (1,1) rectangle (10.5,5.3) ; 
    \node (opt) [anchor = south] at (2,4.7) {p2cudf.jar} ; 
\node(p2cudf)[boiterect] at (3.8,2) {cudf2p2};
\node (p2req)[inout] at (3.8,4) {p2 Request};
\node(p2)[boiterect] at (6.5,4) {p2};
\node(sat4j)[boiteround] at (9,4) {Sat4j-PB};

\node(opb)[boite, text width=2cm] at (6,6) {OPB Problem};
\node(opb-map)[boite, text width=2cm] at (9,6) {OPB Mapping};

\draw [->] (cudf) -- node  [anchor = south] {1} (p2cudf) ;
\draw [->,blue] (p2) -- (opb-map);
\draw [->,blue] (p2) -- (opb);
\draw [->,green] (p2cudf) -- node  [anchor = east] {2}(p2req) ;
\draw [->,green] (p2req) -- node  [anchor = south] {3} (p2);
\draw [<->,red] (p2) -- node  [anchor = south] {4} (sat4j) ;
\draw [->,green] (p2) -- node  [anchor = south] {5} (p2cudf) ;
\draw [->] (p2cudf) -- node  [anchor = south] {6} (cudfsol) ;
\end{tikzpicture}
\end{center}
\caption{Flow diagram of p2cudf to get a solution from a CUDF instance.}
\label{flow-diagram-p2cudf}
\end{figure}
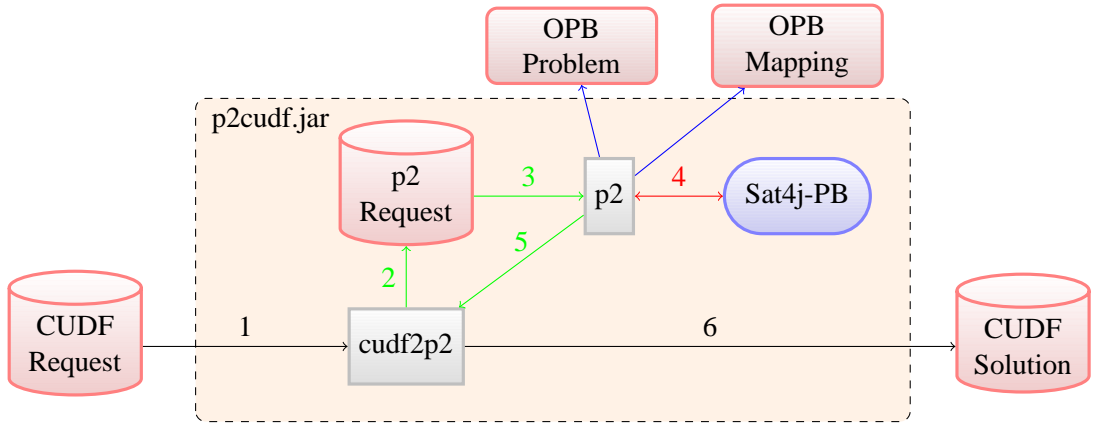
The aim of the P2CUDF resolver was to use the software p2\cite{lbriwoce2009} developed for the Eclipse platform in the context of the Linux distribution to see if such approach was reusable in another context. The software developed for Eclipse could be reused ``as is'' since there are just a few differences between CUDF and Eclipse metadata, the main ones being the use of ranges in the version of the packages, and the notion of optional and non greedy dependencies between packages (see \cite{lbriwoce2009} for details). However, in order to allow the integration of the various criteria defined by the Mancoosi project, a simplified version of p2 code was used. The organization of the solver is depicted in Figure \ref{flow-diagram-p2cudf}.

The first step is to translate the CUDF request into a p2 request: each CUDF package is translated into an Installable Unit\cite{lbriwoce2009}. Once the translation is finished, the p2 resolver translates the request into a pseudo-Boolean optimization problem. It can either directly feed Sat4j-PB solver or output it in a file with the corresponding mapping. In the former case, the p2 resolver gets a solution from Sat4j-PB. That solution is then translated into a simplified CUDF universe (with only the package name, version and installation status).

The next sections describe more formally the constraints used in p2 and the way the optimization functions have been designed to implement the Mancoosi optimization criteria.
\subsection{Original constraints}
\label{p2cudf:constraints}
In the following, $p_{i}^{v}$ will denote package $p_{i}$ in version $v$. We will use the same notation to represent the propositional variables. We will simply write $p_{i}$ when no information is provided for the version. $versions(p_{i})$ will denote the set of available versions of the package $p_{i}$ in the universe. $Installed$ and $NonInstalled$ will represent the versions of the packages that are respectively installed or not in the universe.\\

CUDF constraints are translated as follows:
\begin{itemize}
\item $p_{i}^{x}$ depends on $p_{j}^{x_{j}}, j \in [1..n]$ is translated into a clause $\lnot p_{i}^{x} \vee p_{1}^{x_{1}} \vee ...  \vee p_{n}^{x_{n}} $
\item $p_{i}^{x}$ conflicts with $p_{j}^{x_{j}}$ ($i \not = j$) is translated into the binary clause $\lnot p_{i}^{x} \lor \lnot p_{j}^{x_{j}}$
\item In case a specific package $p_{i}^x$ conflicts with $p_{i}$, which means that no more than one version of the same package can be installed, we use a cardinality constraint $\sum_{x \in versions(p_{i})} p_{i}^{x} \leq 1$
\item A package $p_{i}^{v}$ that cannot be found in the universe will be denoted by the negative literal $\lnot p_{i}^{v}$
\item A package $p_{i}^{v}$ that is requested to be installed or updated will be denoted by the positive literal $p_{i}^{v}$
\end{itemize}

\subsection{Handling inconsistency of installed packages}
\label{p2cudf:inconsistency}

One of the big differences between Eclipse metadata and Linux metadata is that an Eclipse profile (the current installation) is always consistent with the metadata while a Linux installation might not be consistent with the metadata (the user may force the installation of a package). This is partly due to the fact that in the Linux world, dependencies are meant to describe what has been validated by QA: they denote preferred configurations, i.e. violating those constraints may still end up with a runnable system. In the Eclipse ecosystem, the dependencies describe the requirements of the classloader:  if those dependencies cannot be satisfied, then the dependent package will not be activated. 

As a consequence, when a CUDF universe denotes a set of installed packages, those packages dependencies might not be satisfied: mapping each installed package $p_{i}^{v}$ to the positive literal $p_{i}^{v}$ -- as it is the case in the Eclipse PBO encoding -- would end up in that case with an inconsistent pseudo-Boolean optimization problem.

For that reason, the packages marked as ``installed'' in the CUDF universe are considered as ``optional requirements'' in the Eclipse p2 terminology: the solver will try to install as many of them as possible, but will not fail if none of them can be installed.

This is achieved by the following constraints, for which we introduce a $Noop$ propositional variable that will prevent the clause to be falsified if none of optional requirements can be installed:
$$Root \rightarrow p_{1}^{v_{1}} \lor p_{2}^{v_{2}} \lor ... \lor p_{n}^{v_{n}} \lor Noop\mbox{ for }p_{i}^{v_{i}} \in Installed$$
$$Noop \rightarrow \lnot p_{i}^{v_{i}}\mbox{ for }i\in [1..n]$$
The fact that a maximum of $p_{i}^{v_{i}} \in Installed$ will be installed will be managed by the optimization function.
\subsection{Objective functions}
The most important part of the translation is to correctly express the criteria used in the competition. We are using here a translation of the lexicographic preference into a single optimization function. This is possible because our pseudo-Boolean engine does support arbitrary precision coefficients. 
For each measure used in a criterion, we introduce new variables and constraints between those new variables and existing ones in order to express the optimization criteria only on newly introduced variables. 
\subsubsection{Common definitions}
A package is removed in the solution if it was installed but is no longer available in any version.
For any package $p_{i}$ installed in the original CUDF, we introduce a new variable $removed_{p_{i}}$ such that $$removed_{p_{i}} \equiv \bigwedge_{p_{i}^{x} \in versions(p_{i})} \lnot p_{i}^{x}, p_{i} \in Installed$$

A package is changed in the solution if the status of one of its versions (installed or not) has changed.
For any package $p_{i}$ in the original problem, we introduce a new variable $changed_{p_{i}}$ that is true {\em iff} the status of any version of $p_{i}$ has changed:$$changed_{p_{i}} \equiv \bigvee_{p_{i}^{x} \in Installed \cap Versions(p_{i})} \lnot p_{i}^{x} \bigvee_{p_{i}^{x} \in NonIntalled \cap Versions(p_{i})} p_{i}^{x},  p_{i} \in U$$

A package is not up to date if that package is installed but the latest version available is not installed. For any package $p_{i}$ in the original problem, we introduce a new variable $not uptodate_{p_{i}}$ such that $$notuptodate_{p_{i}} \equiv  \lnot latest(p_{i}) \land (\bigvee_{p_{i}^{x} \in versions(p_{i})\setminus latest(p_{i})} p_{i}^{x}), p_{i}  \in U$$.

A package is new if the package was not installed but appears installed in the solution. For any package $p_{i}$ that was not installed in the original CUDF, we introduce a new variable $new_{p_{i}}$ such that $$new_{p_{i}} \equiv \bigvee_{p_{i}^{x} \in versions(p_{i})} p_{i}^{x}, p_{i} \in NonInstalled$$
\subsubsection{Paranoid criterion}
The paranoid criterion is a lexicographic preference on the number of packages removed and then on the number of changed packages.
In that context, our optimization function on  $|Installed| + |U|$ variables is $$ (|U|+1) \times \sum removed_{p_{i}} +  \sum changed_{p_{j}}$$
\subsubsection{Trendy criterion}
The trendy criterion is a lexicographic preference on the number of packages removed, the number of packages that are not up-to-date and the number of newly introduced packages. 
In that context, our optimization function on $ 2 \times |U|$ variables is $$ (|U|+1) \times (|U|+1) \times  \sum removed_{p_{i}} +  (|U|+1) \times \sum notuptodate_{p_{j}} +  \sum new_{p_{k}}$$

\section{Multi-Level Optimization through PWMS}

From Figure \ref{flow-diagram-p2cudf}, it is clear that Sat4j-PB can be replaced by another PB solver. Another not so straightforward alternative would be to use a MaxSAT solver instead, which would require an additional step translating a given instance in the OPB format to the Partial Weighted MaxSAT (PWMS) format.

This section describes the use of the MSUnCore MaxSAT solver to get a solution from a CUDF instance. MSUnCore is a MaxSAT solver which is known for being particularly suitable for solving large problem instances coming from real applications. MSUnCore relies on the identification of unsatisfiable subsets of clauses. With this purpose, a SAT solver is called iteratively and relaxation variables are added to the clauses belonging to the unsatisfiable subset of clauses, jointly with at most one constraints for the new variables. At the end, the solution can be obtained from the values given to relaxation variables. Further details can be found in~\cite{jpms-sat09}.

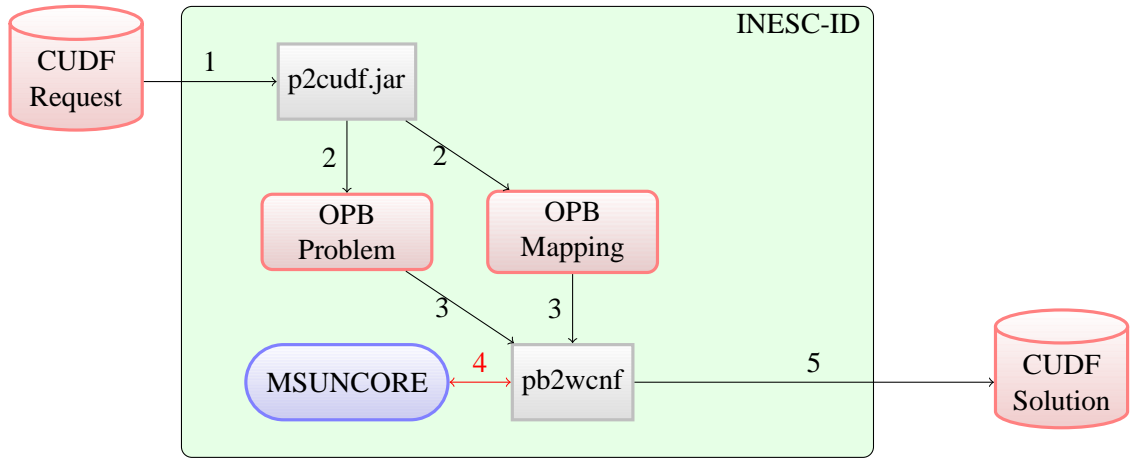
\begin{figure}[t] 
\begin{center}
\begin{tikzpicture}
\node (cudf)[inout] at (-0.6,6) {CUDF Request};
\node (cudfsol)[inout] at (12.5,2) {CUDF Solution};

    \path  [fill=green!10, draw=black, rounded corners] 
    (0.8,1) rectangle (10,7) ; 
    \node (inesc) [anchor = south] at (9,6.5) {INESC-ID} ; 


\node (p2cudfjar)[boiterect] at (3,6)  {p2cudf.jar} ; 

\node(opb)[boite, text width=2cm] at (3,4) {OPB Problem};
\node(opb-map)[boite, text width=2cm] at (6,4) {OPB Mapping};


\node(p2wcnf)[boiterect] at (6,2) {pb2wcnf};
\node(msuncore)[boiteround] at (3,2) {MSUNCORE};


\draw [->] (cudf) -- node  [anchor = south] {1} (p2cudfjar);
\draw [->] (p2cudfjar) -- node  [anchor = east] {2} (opb-map);
\draw [->] (p2cudfjar) -- node  [anchor = east] {2} (opb);
\draw [->] (opb-map) -- node  [anchor = east] {3} (p2wcnf);
\draw [->] (opb) -- node  [anchor = east] {3} (p2wcnf);
\draw [<->,red] (p2wcnf) -- node  [anchor = south] {4}  (msuncore);
\draw [->] (p2wcnf) -- node  [anchor = south] {5} (cudfsol);

\end{tikzpicture}
\end{center}
\caption{Flow diagram of INESC-ID solver to get a solution from a CUDF instance.}
\label{flow-diagram}
\end{figure}

From an implementation point of view, and in order to integrate MSUnCore in the flow we have to take into account the steps depicted in Figure~\ref{flow-diagram}. The core of the flow is called INESC-ID, as the solver has been developed within the participation of INESC-ID in the Mancoosi project. (Note, however, that the solver was also contributed by UdL and UCD, and uses p2cudf as a front-end.) Following the use of p2cudf.jar, we have to translate the OPB file produced by p2cudf.jar to the PWMS format. This translation is done with the tool pb2wcnf. Both tools, p2cudf.jar and pb2wcnf, also produce different mappings between the original source and the variables used in the output encoding. In the case of p2cudf.jar, it writes a file with the mapping between the variables of the OPB format and the package names and versions of the original CUDF file. The tool pb2wcnf produces the mapping between the Boolean variables and the OPB variables. 

The way the different components of the flow have been integrated has several advantages. One of them is that each one of the main components of the process is a standalone tool that can be build independently from the others. Another advantage is that we can easily replace any of the components without affecting the other components. The main drawback of this method is that the communication between tools is currently achieved through files. It makes the implementation easier but can require significant memory and slows down the whole process if the files are too large.

\subsection{Constraints and lexicographic optimization in MaxSAT}

The goal of the tool pb2wcnf is to translate the pseudo-Boolean constraints and the objective function generated by p2cudf to a MaxSAT formula. Actually, the resulting formula corresponds to a Partial Weighted MaxSAT (PWMS) formula, where clauses can be hard or soft and soft clauses are associated with weights (being represented as $(C, w)$). A solution to a PWMS formula satisfies all the hard clauses and maximizes the sum of the weights associated to satisfied soft clauses. Note that in practice all the hard clauses are associated with a weight given by the sum of the weights of all soft clauses plus one.

The translation from PB-constraints to PWMS is performed as follows:

\begin{itemize}
\item The pseudo-Boolean constraints which correspond to clauses are translated to hard clauses. These constraints have been presented in sections~\ref{p2cudf:constraints} and~\ref{p2cudf:inconsistency}. The first correspond to dependencies, conflicts, packages that cannot be found in the universe and packages that are requested to be installed. The latter correspond to handling the inconsistency of installed packages.
\item The remaining pseudo-Boolean constraints, which correspond to the cardinality constraint $\leq 1$, meaning that no more than one version of each package can be installed, are converted to hard clauses as well. In this case, each constraint is encoded into a set of hard clauses using the bitwise encoding~\cite{prestwich-sat07}. Considering that we have $m$ versions of the same package, the bitwise encoding introduces $O(log\;m)$ new variables and $O(m\;log\;m)$ binary clauses.
\end{itemize}

Moreover, the optimization function in the formula generated by p2cudf is translated to a set of weighted soft clauses having only one literal each. The weight associated to each clause is extracted from the coefficients in the optimization function. The translation is therefore performed as follows, depending on the criterion being used:

\begin{itemize}
\item In the paranoid criterion the optimization function is $(|U|+1) \times \sum removed_{p_{i}} +  \sum changed_{p_{j}}$. Hence, for each package $removed_{p_i}$ is created a weighted soft clause $(\neg removed_{p_i},|U|+1)$ and for each package $changed_{p_{j}}$ is created a weighted soft clause $(\neg changed_{p_{j}},1)$.
\item In the trendy criterion the optimization function is $(|U|+1) \times (|U|+1) \times  \sum removed_{p_{i}} +  (|U|+1) \times \sum notuptodate_{p_{j}} +  \sum new_{p_{k}}$. Hence, for each package $removed_{p_i}$ is created a weighted soft clause $(\neg removed_{p_i},(|U|+1) \times (|U|+1))$, for each package $notuptodate_{p_{j}}$ is created a weighted soft clause $(\neg notuptodate_{p_{j}},|U|+1)$ and for each package $new_{p_{k}}$ is created a weighted soft clause $(\neg new_{p_{k}},1)$.
\end{itemize}

The weight of the hard clauses is therefore $(|U|+1) \times (|U|+1)$ for the paranoid criterion and $(|U|+1) \times (|U|+1) \times (|U|+1)$ for the trendy criterion.

\subsection{Lexicographic Optimization with MaxSAT solving}

Once we have translated the original CUDF problem instance to the OPB format, using p2cudf.jar, and the resulting OPB file to PWMS, using pb2wcnf, we can run the MaxSAT solver MSUnCore with the PWMS file obtained. 

The actual version of MSUnCore being used has been enhanced with the possibility of handling lexicographic optimization functions as a sequence of optimization functions. This means that the optimal value for the first optimization criterion is found first, then the optimal value to the second criterion subject to the optimal value to the first criterion, and so on. (The use of this approach both in the context of MaxSAT and PBO is further detailed in~\cite{jpms-ijcai09}.) Not only this approach can be more efficient for solving some problems, but also it takes advantage of the competition evaluation rules which consider that better non-optimal solutions give better values to the most important criteria. 

Let us illustrate a call to MSUnCore for which the optimization function corresponds to the trendy criterion. The soft clauses can therefore be of three different types:  $(\neg removed_{p_i},(|U|+1) \times (|U|+1))$, $(\neg notuptodate_{p_{j}},|U|+1)$ and $(\neg new_{p_{k}},1)$, which requires three iterations:
\begin{enumerate}
\item In the first iteration, MSUnCore finds a solution to the hard
  clauses and the soft clauses with the largest weight (i.e.
  $(|U|+1)\times(|U|+1)$). Since all soft clauses have the same
  weight, the problem being solved is an instance of partial MaxSAT.
  The MinUNSAT solution computed by MSUnCore is denoted $u_1$. In addition,
  by the time the search has finished, MSUnCore has changed the original 
  formula, by adding relaxation variables and at most one constraints as 
  required to solve the problem while identifying unsatisfiable subsets of
  clauses. All clauses in the modified CNF formula are declared {\em hard}
  for the next iteration.
\item In the second iteration, MSUnCore finds a solution to the
  modified set of hard clauses and the soft clauses with the next
  weight (i.e.~$(|U|+1)$). As before, all soft clauses have the same
  weight and so the problem being solved is an instance of partial
  MaxSAT. The new solution computed by MSUnCore is denoted $u_2$. In
  addition, and as before, all clauses in the modified CNF formula
  are declared {\em hard} for the next iteration.
\item Finally, for the third iteration, MSUnCore finds a solution to
  the modified set of hard clauses and the soft clauses with weight
  $1$. Again MSUnCore solves an instance of partial MaxSAT. The new
  solution computed by MSUnCore is denoted $u_3$.
\end{enumerate}
The final solution is given by $u_1$, $u_2$ and $u_3$.
This approach has the advantage of completely eliminating the direct manipulation of weights by the MaxSAT solver. 
In addition, even when the solver has not yet run all the iterations, it is possible to provide a solution which can be already optimized for the first requirements. A more detailed description of this approach is described elsewhere~\cite{jpms-rcra10}.

If MSUnCore ends with a solution within the available CPU time, we can proceed to parse the output of the solver and identify the Boolean variables set to \texttt{true}. Otherwise, if the solver cannot find a solution within the CPU time given, the solver outputs the best solution found so far and we can proceed the same way. A Boolean variable with the \texttt{true} value assigned means that the package referenced by this variable must be installed to get the optimum solution given by MSUnCore. Following the mapping from Boolean variables to OPB variables, and the mapping from OPB variables to package names, the package names and respective versions that have to be installed in the optimal solution can be identified. As a final step, the list of package names with the respective version is output as the CUDF solution.


\section{Experimental Results}

The following experiments were run as part of the Mancoosi internal Solver Competition (MiSC) in January and February 2010. The results presented here are based on the results available online at \url{http://www.mancoosi.org/misc-live/20100208/}.
The solvers were run on different categories of benchmarks: the 10orplus and 9orless categories come from a Debian distribution universe with 45598 packages, the request being solved by apt-get (Debian default resolver) after removing more or less than 10 installed packages. Caixa benchmarks come from Caixa Magica Linux distribution. They contain around 20K packages. Finally, the remaining ones have been randomly generated from a wide Debian universe (including unstable packages). The size of the universe is respectively 51449 packages for biglist and newlist and 31603 packages for smallist. Remember that the size of the universe affects the size of the encoding and the size of the optimization function.

Tables \ref{table-paranoid} and \ref{table-trendy} summarize the results for the three best performing solvers: P2CUDF, INESC-ID and UNSA~\cite{unsa-misc10}. For each category of benchmarks, we provide the number of instances to solve and the number of instances for which no solver could find a solution. This is the case, for instance, when one of the packages to install is not available or one of its dependencies cannot be satisfied. We also provide, for each solver, the number of optimal solutions found and the maximum CPU time required to compute that solution: TO denotes that the timeout (300s) was reached. If an optimal solution cannot be provided within the timeout, we add into parenthesis the number of non-optimal solutions returned by the solver. Table \ref{table-paranoid} provides the results for the paranoid criterion, a lexicographic optimization on two criteria, while Table \ref{table-trendy} provides the results of the trendy criterion, a lexicographic optimization on three criteria.
\begin{table}
\begin{tabular}{|l|r|r|r|r|r|r|r|r|r|}
\hline
\multicolumn{4}{|c|}{} & \multicolumn{2}{|c|}{P2CUDF} &\multicolumn{2}{|c|}{INESC-ID}&\multicolumn{2}{|c|}{UNSA}\\
\hline
Category & \# pkg &\#instances & \#FAIL & Solved & Max T. & Solved & Max. T & Solved & Max. T\\
\hline
10orplus & 45598 &40 & 0 &  0(40) & TO & {\color{blue} 40}& 18s & {\color{blue} 40} & 18s\\
9orless & 45598 &38 & 0 & 0(40) & TO &{\color{blue} 38} & 108s & {\color{blue} 38} & 18s \\
\hline
caixa & 20K &40 & 30 & {\color{blue} 10} &5s & {\color{blue} 10} &6s & {\color{blue} 10} &2s\\
\hline
biglist & 51449&27 & 11 & 9(7) & TO & {\color{blue} 15}(1) & 52s & {\color{blue} 15}(1) & 6s\\
newlist & 51449& 28 & 3 &  {\color{blue} 25} &180s & 24(1) &TO& 24(1) &6s\\
smallist & 31603 &30 & 12 & 17(1) & TO&  {\color{blue} 18} & 10s &  {\color{blue} 18} & 2s  \\
\hline
\end{tabular}
\caption{Results of P2CUDF, INESC-ID and UNSA during the Mancoosi internal Solver Competition, Paranoid criterion, as of Feb 8, 2010}
\label{table-paranoid}
\end{table}

\begin{table}
\begin{tabular}{|l|r|r|r|r|r|r|r|r|r|}
\hline
\multicolumn{4}{|c|}{} & \multicolumn{2}{|c|}{P2CUDF} &\multicolumn{2}{|c|}{INESC-ID}&\multicolumn{2}{|c|}{UNSA}\\
\hline
Category & \# pkg &\#instances & \#FAIL & Solved & Max T. & Solved & Max. T & Solved & Max. T\\
\hline
10orplus & 45598 &40 & 0 & 0(40) & TO& {\color{blue} 40} &92s & 33(7) & 69s\\
9orless & 45598 &38 & 0 & 0(40) & TO & {\color{blue} 38} & 72s & 29(9) & 46s\\
\hline
caixa & 20K &40 & 30 & 6(4) & 5s & {\color{blue} 10}& 6s &7(3)&3s\\
\hline
biglist & 51449&27 & 11 & 8(8) & TO &{\color{blue} 16} & 20s &  {\color{blue} 16} & 9s\\
newlist & 51449& 28 & 3 &  11(14) &TO &  {\color{blue} 25} & 13s&{\color{blue} 25} & 11s\\
smallist & 31603 &30 & 12 & {\color{blue} 18} & 169s & {\color{blue} 18} & 9s&{\color{blue} 18} & 3s\\
\hline
\end{tabular}
\caption{Results of P2CUDF, INESC-ID and UNSA during the Mancoosi internal Solver Competition, Trendy criterion as of Feb 8, 2010}
\label{table-trendy}
\end{table}

Not surprisingly, the time needed to find an optimal answer for the trendy criterion is in general greater than the time needed to find an optimal solution for the paranoid criterion. Furthermore, all of the benchmarks have been solved by at least one solver.
 
One can note that INESC-ID and UNSA solvers outperform P2CUDF. It is interesting to note that one can argue that the main difference between INESC-ID and P2CUDF is simply the back-end resolver: Sat4j-PB\cite{sat4j} for P2CUDF, and MSUnCore\cite{jpms-sat09} for INESC-ID.
On those benchmarks, Sat4j-PB is one order of magnitude slower than MSUnCore. This is not always the case: during the MAXSAT09 evaluation, Sat4j-MAXSAT (based on Sat4j-PB) outperformed MSUnCore on several classes of benchmarks. Better results could be obtained with P2CUDF on those benchmarks after noticing that 30\% of the time is spent on updating the heuristic of the SAT solver because of the huge objective function found on some benchmarks (around 12K literals for 10orplus or 9orless benchmarks). However, we did not fix that first issue because it requires a change in Sat4j that does not pay off in other contexts.
Furthermore, p2 uses by default the assumption based satisfiability proposed in Minisat~\cite{een-sat03}. Propagating unit literals instead (as found in the generated OPB file) allows to improve greatly the running time (because the consequences of propagating that unit clause can be kept during the optimization process, which is not the case using assumption based satisfiability). That second issue has been fixed, but the improvements cannot be fairly reported here because we do not have a comparable hardware to run our own tests.\\

Another interesting fact is that in the few cases where INESC-ID or UNSA are not providing the optimal solution, their behavior is different.
{\em UNSA answers before reaching the timeout:} in that case, it is likely that there is a problem in the encoding of the optimization function or that CPLEX reports an incorrect result due to its use of floating point arithmetic 
\cite{incorrectcplex}. This is especially true for the trendy criterion: most often the optimal solution found by UNSA differs by one package only from the one of INESC-ID in the third measure.
{\em INESC-ID reaches the timeout:} improvements in that case can be achieved by tuning the solver parameters to the specificities of CUDF benchmarks.

Note that because of the way INESC-ID and P2CUDF manage the criterion to optimize, the solutions provided by the solvers are different in case of timeout: 
INESC-ID provides a solution for which the first criterion is almost certainly optimal (unless it got stuck while optimizing the first criterion), but the remaining criteria are usually far from the optimum ;
P2CUDF provides a solution that has been globally optimized, i.e. that should not be far from the optimal for each criterion.

It is interesting to note that on solving the caixa benchmarks using the trendy criterion, P2CUDF and INESC-ID while sharing the same encoding did find different ``optimal'' answers: the implementation of the ``notuptodate'' criterion was invalid in P2CUDF: it was implemented by simply putting a penalty on non latest versions of packages. So if a new version of a package is available, and the previous one is still needed by some dependency but the latest version can safely be installed, P2CUDF and INESC-ID behave differently because nothing in the objective function can guide them in that case: P2CUDF does not install non required packages while INESC-ID does.

\section{Conclusion}
We presented two tools developed in the context of the Mancoosi project that aim at solving Linux upgradeability problems. Both of them share the same translation of the upgradeability problem into a Boolean optimization problem, provided by the tool P2CUDF. INESC-ID is currently one of the best performing solvers on the Mancoosi upgradability problems database, thanks to (i) a correct interpretation of the paranoid and trendy optimization criteria defined by the Mancoosi project and (ii) a very efficient boolean optimization engine (MSUnCore) which implements multi-level optimization.

The next natural step is to participate in the Mancoosi International Solver Competition (MISC). Contrariwise to the internal competition for which no existing solvers or known results were available, some solvers and benchmarks with expected answers are now available so we expect to tune our solvers to obtain better results.
Some existing benchmarks are already a challenge for P2CUDF so we will work on improving the behavior of Sat4j-PB on those particular benchmarks.
One of the main differences between P2CUDF and INESC-ID is the way the lexicographic criteria are handled: INESC-ID optimizes each criteria in turn, following the lexicographic order, while P2CUDF uses a single global optimization function. Considering that UNSA is following an approach similar to INESC-ID, it is tempting to conclude that the iterative approach should be preferred for P2CUDF as well. It would reduce at each step the size of the optimization function, whose size if currently a problem for Sat4j-PB. However, since Sat4j-PB has sometimes some difficulty to prove the optimality of a solution, it might simply get stuck on the first criterion. Further experiments are needed to answer that question.

\vspace{0.3cm}
{\bf Acknowledgements}
This work was partially funded by the European project Mancoosi (FP7-ICT-214898), by FCT (INESC-ID multiannual funding) through the PIDDAC Program funds, by French Ministry of Higher Education and Research,
Nord-Pas de Calais Regional Council and FEDER through the `Contrat de Projets Etat Region (CPER) 2007-2013', and by the Spanish CICYT Projects TIN2007-68005-C04-01/02 and TIN2009-14704-C03-01/02.
\vspace{-0.2cm}
\bibliographystyle{eptcs}
\bibliography{lococo}

\end{document}